\pdfoutput=1
\documentclass{article}

\usepackage{main}

\usepackage[utf8]{inputenc} 
\usepackage[T1]{fontenc}    
\usepackage{hyperref}       
\usepackage{xurl}           
\usepackage{booktabs}       
\usepackage{amsfonts}       
\usepackage{microtype}      
\usepackage{graphicx}

\DeclareUnicodeCharacter{2011}{\mbox{-}}

\hypersetup{
  pdftitle={Decentralized Compute on Untrusted Hardware Using Intel TDX and Encrypted CVMs},
  colorlinks=true,
  linkcolor=blue,
  urlcolor=blue,
  citecolor=blue
}

\title{Decentralized Compute on Untrusted Hardware Using Intel{\normalfont\textregistered} TDX and Encrypted CVMs}

\author{
  Venish Patidar, Dhruv Bindra, Ahmed Darwich, Josh Brown \\
  Manifold Labs Inc. \\
   \And
  Haidong Xia, Sathi Nair \\
  Intel \\
}

\begin{document}
\maketitle

\begin{abstract}
The rapid growth of artificial intelligence workloads has generated an unprecedented demand for secure and scalable compute resources. However, centralized cloud providers continue to dominate both pricing and security models. In an increasingly competitive AI landscape — where the compromise of training data or model weights can confer a significant advantage — there is a critical need for a computing infrastructure that safeguards data at rest, in transit, and in use, while remaining affordable and broadly accessible. Furthermore, existing GPU cluster offerings (e.g., 8xH100s, 8xH200s, 8xB200s) create financial barriers that limit access for organizations, startups, and independent researchers seeking secure, high-performance computing environments.

This paper introduces a decentralized, confidential computing platform that leverages Intel® Trust Domain Extensions (TDX) [1], Intel® Trust Authority (ITA) and NVIDIA Confidential Computing (CC) [2] to establish a distributed ecosystem of fully encrypted Confidential Virtual Machines (CVMs). The proposed architecture incentivizes hardware providers to contribute Intel TDX capable compute resources. Each participating provider is provisioned with a freshly instantiated, uniquely encrypted Ubuntu 24.04 [3] CVM, providing data protection across all stages—at rest, in transit, and in use.

By decentralizing the confidential computing stack and leveraging confidential computing across independently operated nodes, this work demonstrates a viable alternative to traditional cloud-based infrastructures. The proposed system offers enhanced security assurances, transparent cost structures, and democratized access to enterprise-grade secure compute capabilities, paving the way for a more open, secure, and equitable foundation for next-generation AI development.
\end{abstract}


\section{Introduction}
Artificial intelligence (AI) has evolved from a research concept into a driving force of technological progress across industries — from high-accuracy protein structure and genome decoding [4] in healthcare to autonomous vehicles and robotics. The rapid adoption of AI, and in particular the emergence of large language models (LLMs) [5], has intensified the demand for high-performance computing infrastructure. Alongside this growing demand, comes an equally critical requirement: security. In many cases, AI innovations involve sensitive intellectual property or proprietary datasets, and compromise of such information can lead to severe competitive, financial, or even societal consequences.

Although centralized cloud providers fulfill much of the global demand for compute, they continue to control both pricing and trust models [6,7]. Their closed and centralized architectures concentrate market power while introducing security concerns such as insider threats, hypervisor vulnerabilities, and limited transparency into hardware-level guarantees. Decentralization offers a compelling alternative: individuals and organizations can contribute unused compute resources to a global network in exchange for incentives, transforming underutilized hardware into productive infrastructure. Similar to sharing-economy models for idle assets, participants can “bring their own compute,” democratizing access and improving global hardware utilization. However, decentralization does not eliminate the trust problem: it redistributes it. When workloads execute on hardware owned and operated by independent providers, strong assurances are needed to help prevent the provider or external adversaries from accessing, inspecting, or tampering with sensitive data. Establishing such trust becomes the central challenge of decentralized AI infrastructure.

At first glance, encryption might appear to solve this problem — but traditional encryption protects data only at rest or in transit, not during execution. This is where hardware-based confidential computing technologies such as Intel TDX become essential. These technologies establish a root of trust at the CPU level, creating a secure enclave where workloads can run in isolation, even from privileged system software. This capability protects data during execution, significantly reducing a critical security exposure in distributed compute environments.

Recognizing this industry need, hardware manufacturers are increasingly aligning toward confidential computing as a first-class design goal. Intel and NVIDIA have introduced confidential computing features directly into their processors and GPUs, enabling enhanced workload protection that does not rely solely on software-level isolation. While these technologies introduce a small performance overhead due to encryption and attestation, the trade-off yields immense benefits in terms of security, compliance, and verifiable trust.

Despite these technological advancements, confidential computing adoption remains constrained by high cost and centralized infrastructure. TEE offerings from major cloud providers are priced at a premium and provisioned conservatively, limiting accessibility and prohibiting smaller participants from leveraging secure compute at scale.

To address these challenges, Manifold Labs introduced a decentralized confidential computing architecture that democratizes access to secure computational resources while preserving strong confidentiality assurances. The platform leverages Intel TDX and NVIDIA Confidential Computing to provision fully encrypted Confidential Virtual Machines (CVMs) on provider-operated hardware. Each participating provider contributes TDX-capable nodes to the network and receives a uniquely encrypted Ubuntu 24.04 CVM, with per-instance cryptographic uniqueness and fingerprinting derived from the provider’s IP association. These CVMs can be decrypted and launched only after successful remote attestation through Intel’s Key Broker Service (KBS), designed to restrict workload execution to verified, attested environments.

By decentralizing the confidential computing stack, our system offers:
\begin{itemize}
\item Enhanced security assurance enabled by modern hardware-backed attestation and hardware-backed encryption across data states (at rest, in transit, and in use)
\item Transparent, market-driven pricing enabled by open participation from provider nodes and competitive resource allocation.
\item Open access to enterprise-grade secure compute, allowing developers, researchers, and organizations to deploy secure AI workloads without relying on centralized cloud providers.
\end{itemize}

\section{Background}
Manifold Labs operates a decentralized, permissionless compute network where hardware providers can register and contribute their machines in exchange for economic incentives. These machines are later used to execute a wide range of workloads through Manifold Labs, Targon Cloud platform, including GPU rentals, serverless execution, and other computational tasks. By design, the system deliberately operates under a zero-trust model where participating hardware providers remain anonymous and unverified, precluding assumptions about host integrity or intentions.

This model introduces a fundamental challenge: although Targon can schedule workloads and manage resources, the physical hardware remains fully controlled by the provider. Customer workloads—whether they involve sensitive data, proprietary models, or confidential computations—must execute on machines potentially operated by untrusted parties. Consequently, traditional security measures such as disk encryption, TLS, or application-level cryptography are insufficient: data must eventually be decrypted in memory for computation, and a malicious host could inspect memory, tamper with execution, or exfiltrate sensitive information.

To address this challenge, Manifold Labs needed a solution that would allow workloads to run on third-party hardware with strong, verifiable assurances that the code and data remain confidential during execution. Specifically, the platform required a mechanism to verify that each encrypted virtual machine launches in a configuration consistent with how it was provisioned, that the decryption key is only released under verifiable conditions, and that the executing VM can be uniquely identified and continuously attested while running on untrusted hardware.

To meet these requirements, Manifold Labs adopted hardware-based confidential computing. Instead of extending trust to the underlying hardware, confidential computing enables workloads to execute inside hardware-enforced isolated environments—or Trusted Execution Environments (TEEs)—where neither the provider nor the host system can access plaintext code or data. By enforcing isolation at the hardware level, TEEs protect data not only at rest and in transit, but also while actively executing in memory, even in the presence of a potentially malicious operating system, hypervisor, or administrator. To enforce controlled, policy-bound decryption key release, the platform integrates Intel Trust Authority and Key Broker Service (KBS). Under this model, cryptographic material is provisioned to each Confidential Virtual Machine only after successful remote attestation of the TEE environment, binding decryption capability to verified hardware and software measurements and helping to reduce the risk of key exposure outside the intended execution context.

\subsection{Trusted Execution Environments}
A Trusted Execution Environment (TEE) is a hardware-enforced isolated execution context that provides confidentiality and integrity protection for code and data during runtime[10]. TEEs establish trust from the silicon level using a hardware-rooted chain of trust, rather than relying on software-based isolation. They enable three key properties: authenticity of the executed code, integrity of runtime state, and confidentiality of memory contents. The protection is available even when the surrounding system software is untrusted, making TEEs particularly well-suited for decentralized and third-party compute environments.

\subsection{Intel TDX and NVIDIA Confidential Compute}
Intel TDX[1,8] implements TEE principles for virtualized environments by introducing Trust Domains (TDs)—\allowbreak cryptographically isolated virtual machines. In TDX, the hypervisor is removed from the trusted computing base, and memory belonging to a TD is encrypted using hardware-managed keys that are inaccessible to the host. Remote attestation allows external parties to verify the identity, configuration, and integrity of a TD before provisioning sensitive workloads. In practice, most enterprise hardware providers participating in decentralized compute networks deploy Intel Xeon processors compatible with TDX, often paired with NVIDIA Hopper or Blackwell series GPUs, as these combinations provide the best support for high-performance confidential workloads.

NVIDIA Confidential Computing extends TEE capabilities to GPUs through hardware-supported enclaves available in the Hopper and Blackwell architectures. When operating in confidential mode, GPUs establish a hardware root of trust and communicate securely with drivers running inside the CPU TEE. Using PCIe passthrough, multiple GPUs can be directly assigned to a single confidential VM while maintaining encrypted communication and isolation from the host system. This enables protected secure execution of large-scale AI training, inference, and general multi-GPU workloads on untrusted infrastructure.

\subsection{Confidential Virtual Machines (CVMs)}
Building on hardware-based TEEs, Confidential Virtual Machines (CVMs) provide a verifiable and enforceable layer of isolation that extends trust from the CPU and GPU to the entire virtual machine. Traditional virtual machines offer resource isolation but assume a trusted host; the hypervisor and system administrators can inspect guest memory, interfere with execution, or exfiltrate sensitive data, making conventional VMs unsuitable for workloads on untrusted infrastructure. When combined with NVIDIA Confidential Compute, this protection extends to GPUs assigned via PCIe passthrough, providing assurance that both CPU and GPU execution remain confidential.

In Manifold’s permissionless compute network, CVMs enable protected execution of sensitive workloads on hardware provider machines with the goal of protecting models weights, data, and execution state from the hardware owner, providing enhanced cryptographic protection even in adversarial hosting environments.

\subsection{Intel Trust Authority: Independent Verification at Scale}
Intel Trust Authority serves as a zero-trust attestation service that validates platform trust across diverse environments including multi-cloud, sovereign clouds, edge, and on-premises deployments. This service complements confidential computing solutions by providing reliable attestation for Intel Trusted Execution Environments (TEEs), giving customers assurance that their applications and data are properly protected on their chosen platforms. ITA serves customers in healthcare, financial services, and regulated industries, providing:
\begin{itemize}
\item Support for composite attestation scenarios, such as verifying both Intel TDX domains and TEE-enabled NVIDIA GPUs simultaneously
\item Reduced risk of data breaches, IP loss, or regulatory violations Independent verification separate from cloud service providers
\item Sensitive data and AI model protection in third-party cloud environments
\item Auditable evidence supporting the security posture of confidential computing environments.
\end{itemize}

\section{System Architecture}
The proposed decentralized confidential computing platform is organized as a layered architecture that combines hardware-rooted trust (Intel TDX and NVIDIA Confidential Computing), deterministic and uniquely fingerprinted VM provisioning, verifiable attestation workflows, decentralized orchestration, and blockchain-based incentives. The architecture also involves four distinct roles: decentralized hardware providers contributing confidential-compute–capable resources, decentralized validators responsible for continuous attestation and security verification, users submitting workloads requiring strong confidentiality protection, and the Targon platform, which coordinates provisioning, verification, scheduling, and incentives across an otherwise untrusted environment. Together, these components form a protected secure, scalable, and economically open compute network. These roles and mechanisms are implemented through five primary architectural layers.

\subsection{Hardware Selection and Compute Layer}
The foundation of the entire infrastructure relies on selecting hardware capable of supporting hardware-enforced confidential computing, which offers significantly stronger security protection than software-based isolation or traditional process isolation models (as demonstrated by multiple enclave breach case studies). Deploying secure AI workloads in untrusted environments requires both Intel TDX capable CPUs (e.g., Intel 5th/6th Gen Xeon “Emerald Rapids/Granite Rapids”) and NVIDIA GPUs with Confidential Computing support (e.g., Hopper H100/H200, Blackwell B200 series). These configurations are designed to launch every Confidential VM with hardware-rooted memory encryption, a protected execution context, and a cryptographically isolated address space from the host. This design helps prevent the host from introspecting, observing, or modifying user workloads. This configuration is designed to mitigate unauthorized tampering with execution state and the leaking of model weights, datasets, or sensitive intermediate representations.

This hardware-rooted foundation forms the root of trust for the entire decentralized confidential computing architecture. All subsequent layers—VM provisioning, attestation workflows, orchestration, and incentives—build on top of this protected base.

\subsection{Confidential VM Provisioning (Targon Image Gateway)}
Once a hardware provider has acquired compatible hardware, enabled Intel TDX in BIOS, and installed the required kernel, virtualization stack, and GPU components, their node becomes eligible to request a Confidential VM (CVM) from Targon. This request is processed by Targon’s Image Gateway service, which generates, encrypts, and prepares a strongly isolated and attestation-ready virtual machine instance.

\begin{figure}[htbp]
  \centering
  \includegraphics[width=0.9\linewidth]{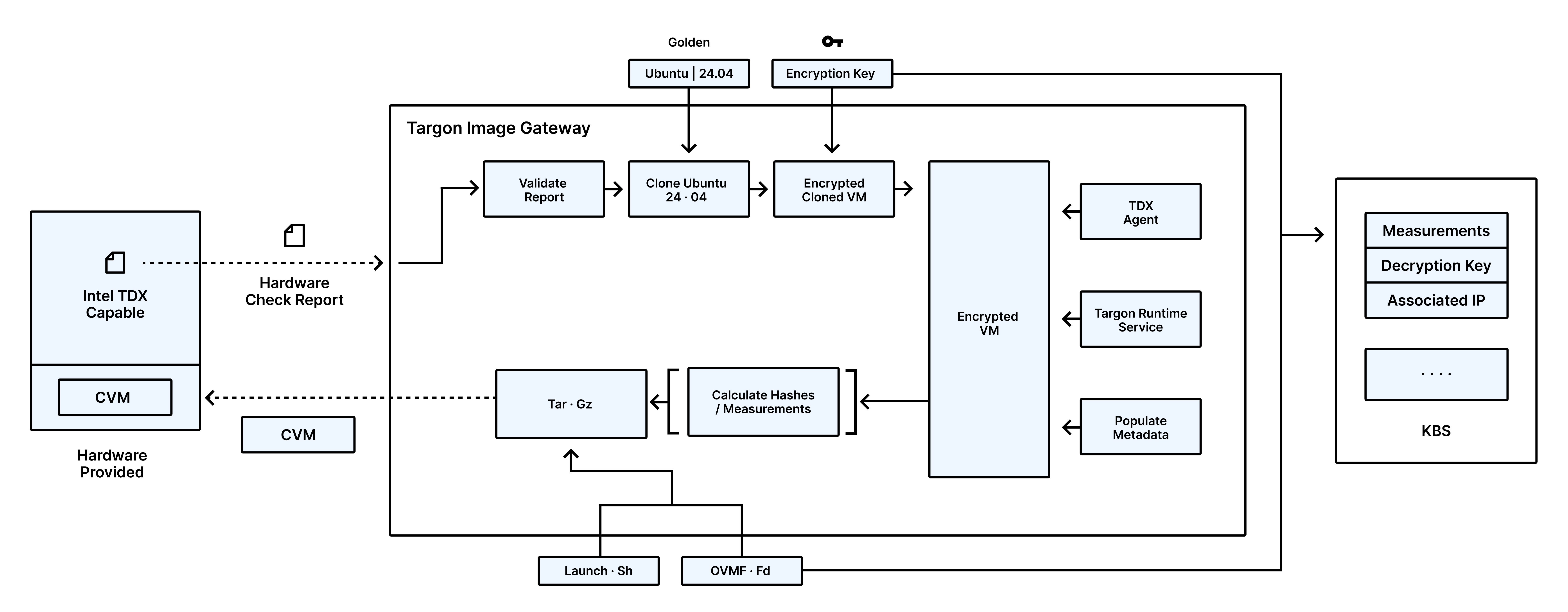}
  \caption{Confidential VM provisioning flow via the Targon Image Gateway.}
  \label{fig:fig1}
\end{figure}

The provisioning pipeline begins by cloning a golden base image: a hardened Ubuntu 24.04 build with pre-installed NVIDIA GPU drivers and confidential-compute-ready configurations. Immediately after cloning, the QCOW2 image is encrypted using a randomly generated per-VM disk key. This key is stored in the Intel ITA Key Broker Service (KBS) and is released only after the VM successfully completes remote attestation. As a result, providers are unable to view, modify, or mount the disk, which protects confidentiality even though they physically own the underlying hardware.

The pipeline then injects all components required for verifiable confidential execution. This includes a Manifold’s Attestation Agent responsible for collecting attestation quote at boot time, sending the quote to ITA for attestation verification and unlocking the disk, as well as the Targon operational runtime services that manage the VM's lifecycle within the decentralized network. As part of this process, the pipeline deterministically computes the expected Intel TDX measurement of the CVM boot chain and records this measurement in the Intel Key Broker Service (KBS), where it is cryptographically associated with the randomly generated per-VM disk decryption key. Finally, the pipeline generates and delivers the launch artifacts required for secure VM startup: a launch script that configures GPUs in Protected PCIe (PPCIe) mode and facilitates proper Intel TDX initialization, along with a validated, TDX-compatible OVMF firmware image.

\subsection{Attestation Workflows}
After the hardware provider successfully requests and downloads the CVM, and then launches it, the VM boots in TDX mode with all attached GPUs operating in PPCIe mode. Before the operating system fully initializes, the Manifold Attestation Agent embedded in the initramfs of the CVM executes the attestation workflow. The agent collects the TDX measurement registers—which contain a cryptographic hash of the entire boot chain, including the kernel, initramfs, and firmware—and generates an attestation quote. This quote is submitted to the Targon Key Broker Service (KBS), which forwards it to the Intel Trust Authority (ITA) for verification [11]. Related work in the context of full‑disk encryption (FDE) also relies on a similar initramfs‑resident agent that retrieves a quote and attests to a remote KBS to securely obtain decryption keys during early boot [12].

\begin{figure}[htbp]
  \centering
  \includegraphics[width=0.9\linewidth]{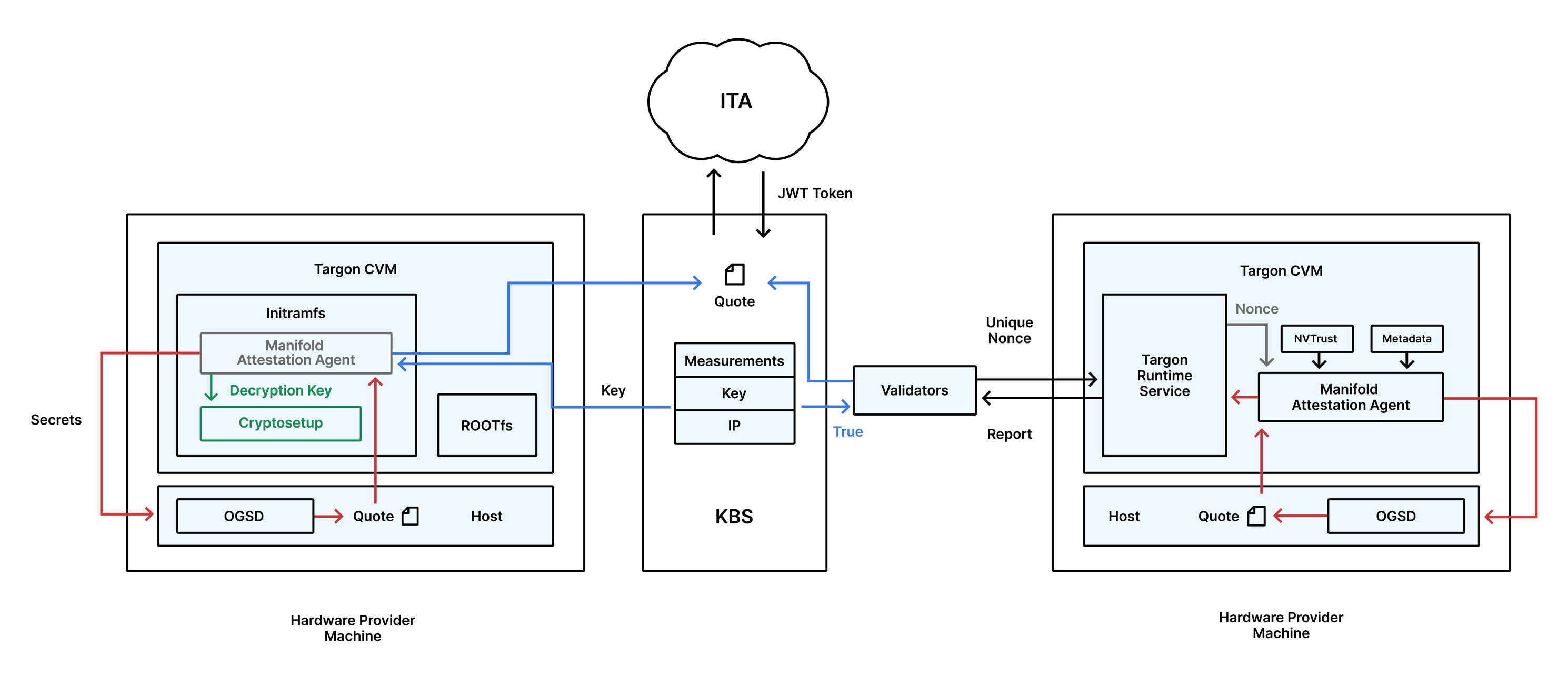}
  \caption{TDX-based early-boot (left) and post-boot (right) attestation workflows for Confidential VMs.}
  \label{fig:fig2}
\end{figure}

Only after attestation succeeds and the reported TDX measurements match the pre-recorded expected values does the Key Broker Service (KBS) release the per-VM disk decryption key. The key is released in a form that is cryptographically bound to the requesting Manifold Attestation Agent and the attested confidential execution context, such that only that specific CVM instance can decrypt and access it. Upon receiving the key, the Manifold Attestation Agent unlocks the encrypted QCOW2 image and allows the boot sequence to proceed. If any component of the boot chain has been modified—whether the kernel, initramfs, firmware, or any other measured element—the TDX measurements will differ from the expected values, attestation will fail, the key will not be released, and the disk will remain encrypted. The system is designed to halt VM launch if attestation fails, leaving the disk encrypted.

While this attestation is performed at boot, the attestation layer also maintains trust across the node's operational lifetime by handling node identity binding, continuous verification of confidential execution, and coordination of attestation across the decentralized validator network. Validators act as a decentralized relay and verification layer, collecting attestation quotes from CVMs, forwarding them to the Key Broker Service (KBS) for Intel/NVIDIA verification, and confirming that each node remains in a valid confidential state. By distributing this responsibility across multiple validators, the network aims to mitigate the risk of any single party manipulating attestation results, providing tamper-resistant, real-time assurance of hardware and VM integrity.

Upon the first successful attestation, the KBS permanently binds the CVM to the hardware provider's network identity by recording the source IP address. All subsequent attestation requests for this VM must originate from the same IP. This binding is designed to mitigate several attack vectors: it makes it significantly more difficult for a hardware provider to copy the encrypted disk to another machine and boot it elsewhere, migrate the VM to a different node, or replay captured attestation flows from a different network location. If a hardware provider's IP address changes for any reason, the existing CVM becomes permanently inaccessible and the provider must request a new VM from the platform.

Beyond the initial boot-time attestation, each CVM performs continuous re-attestation at every block interval (roughly 72 minutes) to prove it remains in a genuine, verified, attested confidential state. To maintain quote freshness and mitigate replay attacks, each attestation round is challenge–response: the validator issues a cryptographically strong nonce, and the CVM includes this nonce in the attestation payload/quote. Validators reject quotes that do not bind the expected nonce, preventing previously captured attestation artifacts from being replayed as valid. This periodic verification helps that the network maintains real-time awareness of each node's security posture rather than relying solely on a one-time boot attestation that could become stale.

At each interval, the Targon runtime services execute a multi-layer attestation workflow. The process begins with NVIDIA nvtrust attestation, which produces a signed report confirming that all attached GPUs are genuine NVIDIA hardware operating in Confidential Computing mode with hardware memory encryption active [13]. The runtime then collects node metadata including GPU specifications, the nvtrust attestation output, and other operational parameters. This combined payload—the NVIDIA attestation report and node metadata—is embedded into the user data field of the Intel TDX quote, effectively nesting the GPU attestation within the CPU attestation called as unified attestation and binding both into a single cryptographic proof.

The generated TDX quote, along with the node metadata, is then submitted to a validator on the network. Because validators do not possess Intel Trust Authority API keys, they forward the attestation package to the Targon Key Broker Service. The KBS extracts the TDX quote and submits it to Intel Trust Authority for verification. Upon successful verification, ITA returns a signed JWT token confirming the quote's authenticity.

The KBS then performs its own validation: it verifies the JWT signature, confirms that the TDX measurements match the expected values for a legitimate CVM boot chain, and validates the NVIDIA attestation report embedded in the user data field. Only if all checks pass is the node marked as live and eligible to receive workload assignments. If any verification step fails—whether the TDX measurements are incorrect, the NVIDIA attestation is invalid, or the JWT cannot be verified—the node is immediately marked as offline.

\subsection{Incentive Layer}
The incentive layer aligns economic motivations with network security and resource availability. Validators not only verify that nodes remain in a valid confidential state, but also facilitate fair compensation for hardware providers based on the compute resources they contribute and the correctness of their attested nodes.

At the start of each interval (approximately every 72 minutes), validators query Targon’s Tower service to retrieve the incentive pool, target nodes, price targets, and caps. The incentive pool defines the total funds available for distribution, while target nodes and price targets determine the nominal reward for different hardware types, such as H100/H200 GPUs or Intel TDX CPU servers.

After completing validation, each validator calculates the contribution of every node relative to the total compute provided and reports these as weights on-chain. The blockchain aggregates the weights across all validators, computing a stake-weighted average for each provider. Payouts are then automatically distributed according to this consensus, ensuring transparency, fairness, and redundancy: multiple validators independently check each node, reducing the risk of manipulation or biased reporting.

By combining continuous attestation with a decentralized, stake-weighted incentive mechanism, the platform encourages honest participation, maintains real-time verification of node integrity, and is designed to ensure fair compensation for hardware providers contributing attested resources.

\subsection{Decentralized Orchestration Layer}
The decentralized orchestration layer is designed to support secure scheduling, execution, and recovery of workloads across a dynamic set of untrusted hardware providers. Participation in this layer is restricted to hardware nodes running CVMs provisioned by the Targon Image Gateway and continuously validated through attestation. By admitting only CVMs in a verified confidential state, the orchestration layer provides an assurance that workloads are executed on attested, verified resources while enabling automated scaling, fault recovery, and efficient utilization of heterogeneous compute assets.

\begin{figure}[htbp]
  \centering
  \includegraphics[width=0.4\linewidth]{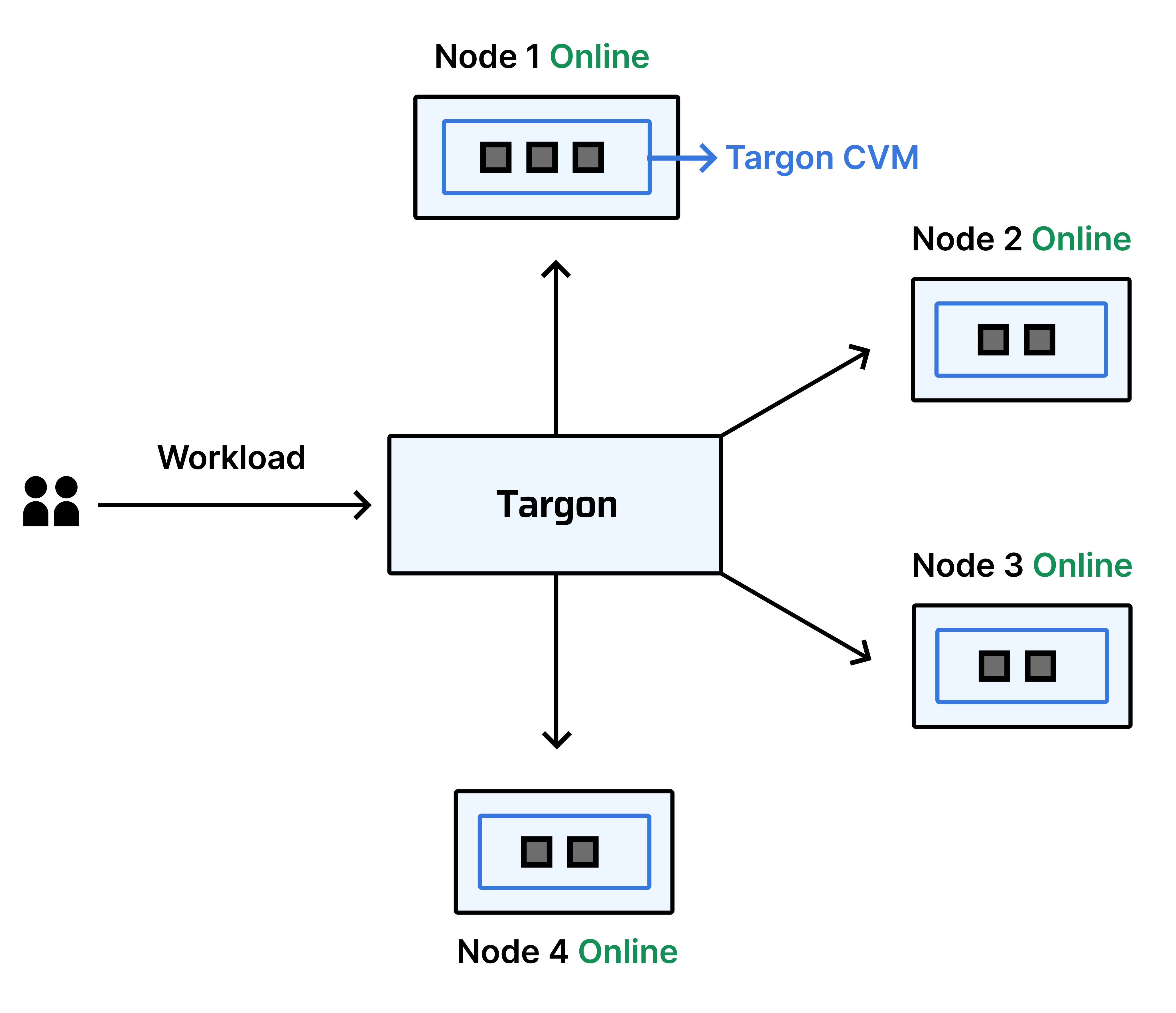}
  \caption{Decentralized orchestration architecture.}
  \label{fig:fig3}
\end{figure}

Once validated, a CVM automatically joins a private WireGuard-based mesh network that interconnects all active nodes. On top of this secure overlay network, Targon operates a Kubernetes control plane, where each CVM registers as a worker node. This design allows the platform to take advantage of the mature Kubernetes’ scheduling, resource management, and failure-handling mechanisms while maintaining strong security protection rooted in confidential computing.

All platform workloads, including GPU rentals, serverless execution, and managed inference services, are deployed as Kubernetes workloads. Only nodes that are in a continuously attested and valid confidential state are admitted into the scheduling pool. As a result, workload placement is restricted to nodes that have been verified to operate in a genuine confidential execution environment. Resource-aware scheduling enables efficient utilization of heterogeneous GPU configurations across the decentralized network.

The orchestration layer is inherently resilient. If a CVM fails continuous attestation, becomes unreachable, or experiences hardware or software failure, it is immediately removed from the scheduling pool. Kubernetes automatically reschedules affected workloads onto other live, attested nodes within the mesh. This maintains both high availability and security, as workload placement dynamically adapts to changes in node health without manual intervention.

By combining confidential computing with a Kubernetes-based decentralized orchestration model, Targon achieves a robust and scalable execution environment that tolerates untrusted infrastructure while providing strong isolation, automated recovery, and seamless workload mobility across hardware provider machines.

\section{Threat Model}
This section defines the adversary model, trust assumptions, and security goals of the system. The design explicitly assumes that hardware providers are untrusted and may behave maliciously.

\subsection{Adversary Model}
The system assumes a strong, adaptive adversary with the following capabilities:
\begin{itemize}
\item Possess full physical control of the host machine.
\item Control the host operating system, hypervisor, BIOS configuration, and firmware settings.
\item Can arbitrarily start, stop, snapshot, or replay virtual machine images.
\item Can observe all network traffic entering and leaving the host.
\item Can attempt to migrate, clone, or replay encrypted VM disks.
\item Can collude with other providers or external entities.
\item Can observe, delay, drop, or replay network messages
\item Cannot break standard cryptographic primitives
\end{itemize}

The adversary is assumed to have no ability to break hardware-backed cryptography, compromise secure key storage within trusted hardware authorities, or forge attestation signatures.

\subsection{Trust Assumptions}
The system relies on the following trusted components:
\begin{itemize}
\item Intel TDX and NVIDIA Confidential Computing are assumed to correctly enforce memory encryption, isolation, and attestation.
\item Side-channel leakage beyond documented vendor threat models is considered out of scope.
\item Intel Trust Authority (ITA) is trusted to verify TDX quotes correctly and to issue authentic verification tokens.
\item NVIDIA GPU attestation mechanisms are trusted to report correct device state.
\item Hash functions, digital signatures, and encryption schemes are assumed to be secure.
\end{itemize}
No trust is placed in:
\begin{itemize}
\item Infrastructure providers
\item Host operating systems or hypervisors
\item Validators
\item Network transport
\end{itemize}

\subsection{Security Goals}
The system is designed to enforce the following security properties:
\begin{enumerate}
\item Confidentiality of workloads: Infrastructure providers must not be able to access CVM memory, disk contents, GPU memory, model weights, datasets, or intermediate computation state.
\item Integrity of execution: Workloads must execute only in configurations that match the provisioned and attested boot state.
\item Authenticity of the execution environment: Workloads must execute only on genuine TDX-capable CPUs and Confidential Computing-enabled GPUs.
\item Non-migratability: Encrypted VM disks must not be reusable on alternate hardware or network identities.
\item Continuous trust enforcement: Nodes that deviate from a valid confidential execution state must be detected and removed.
\end{enumerate}

\subsection{In-Scope Attacks and Mitigations}
\begin{table}[htbp]
  \centering
  \begin{tabular}{@{}p{0.32\linewidth}p{0.62\linewidth}@{}}
    \toprule
    Attack Vector & Mitigation \\
    \midrule
    Disk inspection or modification & Per-VM disk encryption; decryption keys released only after successful attestation \\
    VM replay on different hardware & IP-based node binding enforced by the Key Broker Service (KBS) \\
    Boot-chain tampering & Verification of TDX measurement registers during attestation \\
    GPU memory inspection & NVIDIA Confidential Computing with Protected PCIe (PPCIe) \\
    Attestation replay & Fresh attestation quotes using validator nonces (challenge-response) and continuous re-attestation \\
    \bottomrule
  \end{tabular}
\end{table}

\subsection{Out-of-Scope Threats}
The following threats are explicitly considered out of scope:
\begin{itemize}
\item Compromise of Intel Trust Authority or NVIDIA root signing keys.
\item Microarchitectural attacks not covered by Intel TDX or NVIDIA Confidential Computing threat models.
\item Denial-of-service attacks by infrastructure providers (e.g., powering off hardware).
\end{itemize}

\section{Conclusion}
In this work, we have proposed and implemented a decentralized confidential computing architecture leveraging Intel TDX, a third-party Intel Trust Authority independent attestation service, and NVIDIA Confidential Computing capabilities. The platform enables permissionless, incentivized utilization of otherwise idle global hardware, creating a self-sustaining ecosystem where hardware providers, validators, and users interact securely and transparently. By decentralizing compute resources that were previously concentrated among a few major cloud providers, the system fosters greater competition, resilience, and democratization of high-performance computing.

Importantly, this architecture from Manifold Labs makes premium, enterprise-grade compute accessible at a fraction of the traditional cost. Researchers, startups, small enterprises, and independent developers---who are often unable to afford conventional high-end cloud offerings---can now deploy protected, confidential workloads on high-performance hardware. The platform is live and in production, and is publicly available at Targon\footnote{\url{https://targon.com}}. By combining hardware-rooted confidentiality, verifiable attestation, decentralized orchestration, and economic incentives, the proposed platform demonstrates a practical and scalable model for secure, decentralized AI and high-performance computing, paving the way for a more open, equitable, and sustainable computational landscape. Future work will explore a user-facing approach for independently verifying the confidential execution state of CVMs using Intel Trust Authority attestation services.


\begin{thebibliography}{13}

  \bibitem{ref1}
Intel Corporation. \emph{Intel\textregistered{} Trust Domain Extensions (Intel\textregistered{} TDX) Whitepaper}. \url{https://www.intel.com/content/dam/develop/external/us/en/documents/tdx-whitepaper-final9-17.pdf}. Accessed: 2026-01-25.

\bibitem{ref2}
NVIDIA. \emph{NVIDIA Confidential Computing}. \url{https://www.nvidia.com/en-us/data-center/solutions/confidential-computing/}. Accessed: 2026-01-25.

\bibitem{ref3}
Canonical. \emph{Ubuntu 24.04 LTS}. \url{https://releases.ubuntu.com/24.04/}. Accessed: 2026-01-25.

\bibitem{ref4}
Google DeepMind. \emph{AlphaGenome: AI for Better Understanding the Genome}. \url{https://deepmind.google/blog/alphagenome-ai-for-better-understanding-the-genome/}, 2025.

\bibitem{ref5}
Hugo Touvron et al. LLaMA: Open and efficient foundation language models. \emph{arXiv preprint} arXiv:2302.13971, 2023.

\bibitem{ref6}
Effects of cloud market concentration. \url{https://www.cloudgovernance.org/cloud-governance-issues/effects-of-cloud-market-concentration/}, 2026. Accessed: 2026-01-25.

\bibitem{ref7}
Electro IQ. \emph{Cloud Computing Statistics: Market Size, Adoption \& ROI (2025)}. \url{https://electroiq.com/stats/cloud-computing-statistics/}, 2025.

\bibitem{ref8}
Intel Corporation. \emph{Intel\textregistered{} Trust Domain Extensions (Intel\textregistered{} TDX) Overview}. \url{https://www.intel.com/content/www/us/en/developer/tools/trust-domain-extensions/overview.html}, 2026.

\bibitem{ref9}
Google Cloud. \emph{Confidential VMs}. \url{https://cloud.google.com/confidential-computing/}. Accessed: 2026-01-25.

\bibitem{ref10}
Francesco Conti et al. A survey of trusted execution environments: Foundations, applications, and challenges. \emph{ACM Computing Surveys}, 51(6):1--38, 2019.

\bibitem{ref11}
Intel Corporation. \emph{Intel\textregistered{} Trust Authority --- Independent Attestation for Confidential Computing}. \url{https://www.intel.com/content/www/us/en/security/trust-authority.html}, 2026.

\bibitem{ref12}
Intel Corporation. \emph{Full Disk Encryption with Intel\textregistered{} Trust Domain Extensions}. \url{https://www.intel.com/content/www/us/en/developer/articles/technical/disk-encryption-intel-trust-domain-trust-authority.html}, 2026.

\bibitem{ref13}
NVIDIA. \emph{NVIDIA Attestation Suite and nvTrust Documentation}. \url{https://docs.nvidia.com/attestation/index.html}, 2026.

\end{thebibliography}
\end{document}